\documentclass[10pt,conference]{IEEEtran}

\usepackage{blindtext,graphicx}
\usepackage{times}
\IEEEoverridecommandlockouts
\usepackage[nospace]{cite}
\usepackage{amsmath,amssymb,amsfonts}

\usepackage{textcomp}
\usepackage{setspace}
\def\smallerspacecaption{\vspace{-2mm}}

\usepackage{floatrow}
\usepackage[caption=false]{subfig}
\usepackage{algorithm}
\usepackage[noend]{algpseudocode}
\usepackage[table, dvipsnames]{xcolor}
\usepackage{fancyhdr}
\usepackage{graphicx}
\usepackage{url}
\usepackage{verbatim}
\usepackage{multirow}
\usepackage{hhline}
\usepackage[us]{datetime}
\usepackage{paralist}
\usepackage{color}
\usepackage{soul}
\usepackage{enumitem}
\usepackage[font=footnotesize,labelsep=period,justification=centering]{caption}
\usepackage[implicit=false,hidelinks]{hyperref}
\usepackage{stfloats}
\usepackage{pifont}

\newcommand{\isempty}{{\fontfamily{lmtt}\selectfont IsEmpty}}
\newcommand{\gnn}{{\fontfamily{lmtt}\selectfont GNN}}
\newcommand{\getpredictions}{{\fontfamily{lmtt}\selectfont GET\_PREDICTIONS}}
\newcommand{\sample}{{\fontfamily{lmtt}\selectfont SAMPLE}}
\newcommand{\getkey}{{\fontfamily{lmtt}\selectfont GET\_KEY}}
\newcommand{\untangle}{{\fontfamily{lmtt}\selectfont UNTANGLE}}

\graphicspath{{figures/}}

\usepackage{fancyhdr}

\newcommand{\blue}[1]{\textcolor{black}{#1}}

\usepackage{soul}
\newcommand{\drop}[1]{\textcolor{red}{#1}}
\renewcommand{\drop}[1]{}
\usepackage{circledsteps}
\pgfkeys{/csteps/inner color=white}
\pgfkeys{/csteps/outer color=none}
\pgfkeys{/csteps/fill color=black}
\newdimen\arrayruleHwidth
\makeatletter
\def\Hline{\noalign{\ifnum0=`}\fi\hrule \@height \arrayruleHwidth
\futurelet \@tempa\@xhline}
\makeatother

\makeatletter
\def\blfootnote{\xdef\@thefnmark{}\@footnotetext}
\makeatother

\shortdate
\settimeformat{ampmtime}

\makeatletter
\newcommand*\tabsize{%
	   \@setfontsize\tabsize{6}{7.2}%
}
\makeatother
\algrenewcommand\algorithmicrequire{\textbf{Input:}}
\algrenewcommand\algorithmicensure{\textbf{Output:}}
\renewcommand{\arraystretch}{1.03}

\renewcommand{\IEEEbibitemsep}{0.3pt}
\makeatletter
\IEEEtriggercmd{\reset@font\normalfont\fontsize{6.8pt}{7.11pt}\selectfont}
\makeatother
\IEEEtriggeratref{1}

\makeatletter
\renewcommand\footnoterule{%
  \kern-3\p@
  \hrule\@width 0.5\columnwidth
  \kern2.6\p@}
  \makeatother
\usepackage{amsmath,amsfonts,bm}

\def\mA{{\bm{A}}}
\def\ma{{\bm{a}}}
\def\mz{{\bm{z}}}

\def\mD{{\bm{D}}}

\def\mI{{\bm{I}}}

\def\mW{{\bm{W}}}
\def\mX{{\bm{X}}}

\def\mZ{{\bm{Z}}}

\DeclareMathAlphabet{\mathsfit}{\encodingdefault}{\sfdefault}{m}{sl}
\SetMathAlphabet{\mathsfit}{bold}{\encodingdefault}{\sfdefault}{bx}{n}

\def\gG{{\mathcal{G}}}

\def\gN{{\mathcal{N}}}

\newcommand{\R}{\mathbb{R}}

\begin{document}

\title{\huge \untangle: Unlocking Routing and Logic Obfuscation Using Graph Neural Networks-based Link Prediction}

\author{Lilas~Alrahis$^\ddag$, Satwik~Patnaik$^\dag$, 
Muhammad~Abdullah~Hanif$^\S$, Muhammad~Shafique$^\ddag$, and Ozgur~Sinanoglu$^\ddag$\\[1ex]
$^\ddag$Division of Engineering, New York University Abu Dhabi, UAE\\
$^\dag$Electrical \& Computer Engineering, Texas A\&M University, College Station, Texas, USA\\
$^\S$Institute of Computer Engineering, Technische Universität Wien, Vienna, Austria\\
\normalsize{\{lma387, muhammad.shafique, ozgursin\}@nyu.edu},
\normalsize{satwik.patnaik@tamu.edu},
\normalsize{muhammad.hanif@tuwien.ac.at}
}

\maketitle

\renewcommand{\headrulewidth}{0.0pt}
\thispagestyle{fancy}
\lhead{}
\rhead{}
\chead{\copyright~2021 IEEE.
This is the author's version of the work.
The definitive Version of Record is published in
2021 International Conference On
Computer-Aided Design (ICCAD)}
\cfoot{}

\begin{abstract}
Logic locking aims to prevent intellectual property (IP) piracy and unauthorized overproduction of integrated circuits (ICs). 
However, initial logic locking techniques were vulnerable to the Boolean satisfiability (SAT)-based attacks. 
In response, researchers proposed various SAT-resistant locking techniques such as point function-based locking and symmetric interconnection (SAT-hard) obfuscation.
We focus on the latter since point function-based locking suffers from various structural vulnerabilities. 
The SAT-hard logic locking technique, InterLock~\cite{InterLock}, achieves a unified logic and routing obfuscation that thwarts state-of-the-art attacks on logic locking.
In this work, we propose a novel link prediction-based attack, {\untangle}, that successfully breaks InterLock in an oracle-less setting without having access to an activated IC (oracle). 
Since InterLock hides selected timing paths in key-controlled routing blocks, {\untangle} reveals the gates and interconnections hidden in the routing blocks upon formulating this task as a link prediction problem. 
The intuition behind our approach is that ICs contain a large amount of repetition and reuse cores. 
Hence, {\untangle} can infer the hidden timing paths by learning the composition of gates in the observed locked netlist or a circuit library leveraging graph neural networks.
We show that circuits withstanding SAT-based and other attacks can be unlocked in seconds with $100\%$ precision using {\untangle} in an oracle-less setting.
{\untangle} is a generic attack platform (which we also open source~\cite{webinterface}) that applies to multiplexer (MUX)-based obfuscation, as demonstrated through our experiments on ISCAS-85 and ITC-99 benchmarks locked using InterLock and random MUX-based locking.
\end{abstract}

\begin{IEEEkeywords}
Logic locking,
Routing obfuscation,
Link prediction,
Oracle-less attacks,
Graph neural networks.
\end{IEEEkeywords}

\section{Introduction}
\label{sec:introduction}

The globalization of the integrated circuit (IC) supply chain has led design companies to outsource the fabrication of chips to off-shore, untrustworthy foundries.
Attackers present in these foundries can either steal the design intellectual property (IP) or engage in unauthorized overproduction of ICs~\cite{rostami2014primer}.
The research community proposed various countermeasures such as logic locking, state-space obfuscation, and split manufacturing (amongst others) to ward off such threats. 
Logic locking is a holistic technique that can protect the design IP from untrusted entities (foundry, test facility, and end-user) in the IC supply chain.
Logic locking accomplishes design IP protection by embedding key-controlled logic (key-gates) driven by an on-chip tamper-proof memory~\cite{epic_journal}.
Applying the correct key (known to the designer) unlocks the chip resulting in the correct functionality, whereas the incorrect key results in an incorrect functionality.
Researchers have developed a series of defenses~\cite{epic_journal,JV-Tcomp-2013,cyclic_obfuscation,yasin_SARLock_host_2016,yasin_CCS_2017,delay_locking,fulllock,InterLock,unsail} and attacks~\cite{JV_DAC_2012,Subramanyan_host_2015,cycsat,scansat_aspdac,scansat_journal,nngsat,alrahis2020gnnunlock,shamsi2017appsat,Azar_Kamali_Homayoun_Sasan_2018,yasin_TETC_2017,chakraborty2018sail,alaql2019sweep,li2019piercing,alrahis2019functional}
over the last decade towards enhancing the security of logic locking.
Most notably, Subramanyan \textit{et al.}~\cite{Subramanyan_host_2015} proposed the \textit{Boolean satisfiability} (SAT)-based attack,  which broke all prior locking techniques.
Researchers developed various SAT-resistant logic locking solutions (further details in Sec.~\ref{sec:background}) to defend against the SAT-based attack.
However, with each developed defense, new attack techniques exposed implementation vulnerabilities.

\textbf{In this work}, we focus on one of the most prominent SAT-resistant techniques that thwart\blue{s} the SAT-based attack~\cite{Subramanyan_host_2015} by constructing symmetric interconnection (routing obfuscation).
Such an approach increases the depth of the SAT search tree, ensuring SAT-hard calls~\cite{fulllock}.
Although na\"ive routing obfuscation thwarts the SAT-based attack, it is vulnerable to re-modeling/encoding attacks~\cite{InterLock,sweeney_model}. 
Recently, Kamali \textit{et al.}~\cite{InterLock} proposed \textit{InterLock} to mitigate the drawbacks of the prior locking techniques.
In InterLock, key-controlled routing blocks (KeyRBs) perform routing and logic obfuscation, twisting logic with routing, thereby thwarting state-of-the-art attacks \blue{on} logic locking.
Next, we discuss the challenges as to why there has been no successful attack on InterLock.

\subsection{Key Research Challenges Targeted in this Work}

\begin{figure}[t]
\centering
\includegraphics[width=\textwidth]{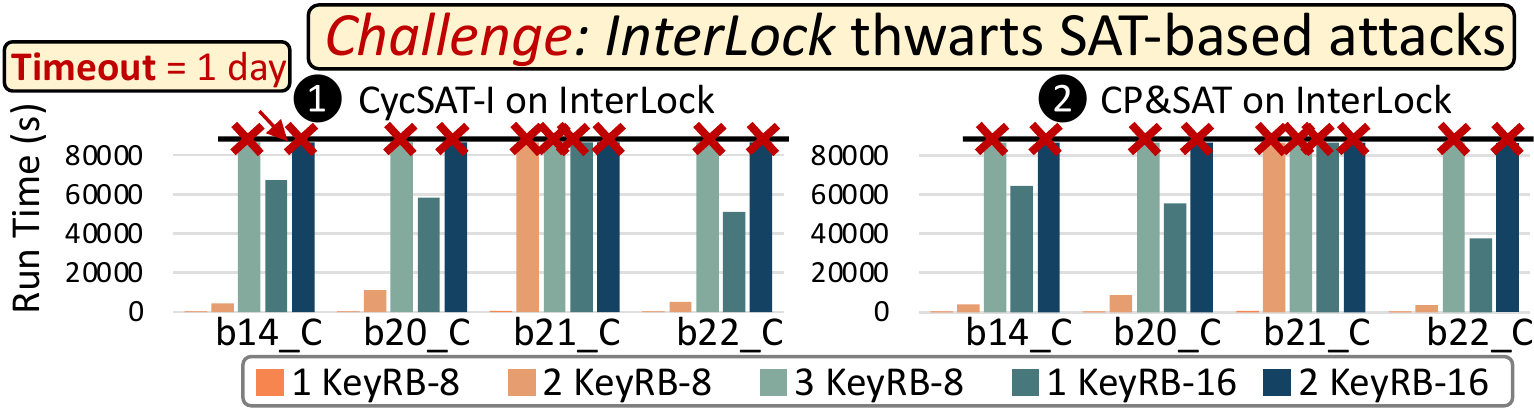}
\caption{The SAT-based CycSAT-I~\cite{cycsat} and CP\&SAT~\cite{InterLock} attacks reach a timeout of one day on benchmarks locked using InterLock with different sizes of key-controlled routing blocks (KeyRBs) (based on the results in~\cite{InterLock}).}
\label{fig:SAT}
\vspace*{-1mm}
\end{figure}

\begin{enumerate}

\item \textit{SAT-hard calls:} InterLock ensures that any attack relying on SAT solvers (e.g., SAT-based attack, \textit{AppSAT}~\cite{shamsi2017appsat}) encounters a complex SAT search tree. 
The authors in~\cite{InterLock} launched the \textit{cyclic-based SAT} (CycSAT-I~\cite{cycsat}) (see~\Circled{\scriptsize\textbf{1}} in Fig.~\ref{fig:SAT}) and the \textit{canonical prune and SAT} (CP\&SAT~\cite{InterLock}) (see~\Circled{\scriptsize\textbf{2}} in Fig.~\ref{fig:SAT}) attacks on locked benchmarks and demonstrated that both attacks fail to recover the secret key, running for a day without termination (timeout).

\item \textit{Multiplexer (MUX)-based locking:} The construction of InterLock utilizes deep MUX trees for locking. 
In general, a MUX key-gate takes an original (true) wire and another (false) wire from the design. 
The select line of the MUX acts as the key-input.
Applying the correct key-bit passes the true wire maintaining the original functionality.
The correct key-bit can either be $0$ or $1$, depending on whether the true wire is connected to the first or the second input of the MUX. 
Hence, an attacker cannot infer the correct key-bit from the type of key-gate, unlike X(N)OR-based locking, which can be broken using machine learning (ML)-based structural attacks\blue{~\cite{chakraborty2018sail,sisejkovic2021challenging}}. 
We illustrate an example of MUX-based locking and showcase the associated challenge in Fig.~\ref{fig:mux_lock}.

\item \textit{Loop formation:} MUX-based locking may introduce combinational cycles in the locked design. 
Note that the SAT-based attack applies only to directed acyclic graphs (DAG)~\cite{Subramanyan_host_2015}.
As a result, cycles trap the attack algorithm in infinite loops.
\blue{Thus, authors in~\cite{InterLock} use the CycSAT-I attack~\cite{cycsat}, which can decrypt cyclic logic encryption, to evaluate the security of InterLock.} 
Although CycSAT-I can handle loops, it faces SAT-hard computations. 
Researchers formulated specific techniques, such as \textit{SWEEP} \blue{and \textit{SCOPE}}, to tackle MUX-based locking~\cite{alaql2019sweep,SCOPE}.
However, \blue{both} SWEEP and \blue{SCOPE} cannot handle cycles in the design.
To demonstrate this key limitation, we lock selected ITC-99 benchmarks using InterLock with 1~KeyRB-16.\footnote{We provide further details about KeyRB construction in Sec.~\ref{sec:background}.}
We also lock selected ISCAS-85 and ITC-99 benchmarks using 2-input MUX-based locking with key-sizes (K) of $\{64, 128, 256, 512\}$, resulting in $24$ locked designs. 
\textit{We observe that \blue{both} SWEEP and \blue{SCOPE attacks fail} to decipher the \blue{keys} due to the presence of loops in the locked designs.}\footnote{SWEEP and \blue{SCOPE rely} on \textit{ABC}~\cite{brayton2010abc} to convert a locked design into a DAG.
When reading a design with combinational loops, the tool reports an error ``Network contains a combinational loop'' and cannot launch the attack.}

\begin{figure}[t]
\centering
\includegraphics[width=0.9\textwidth]{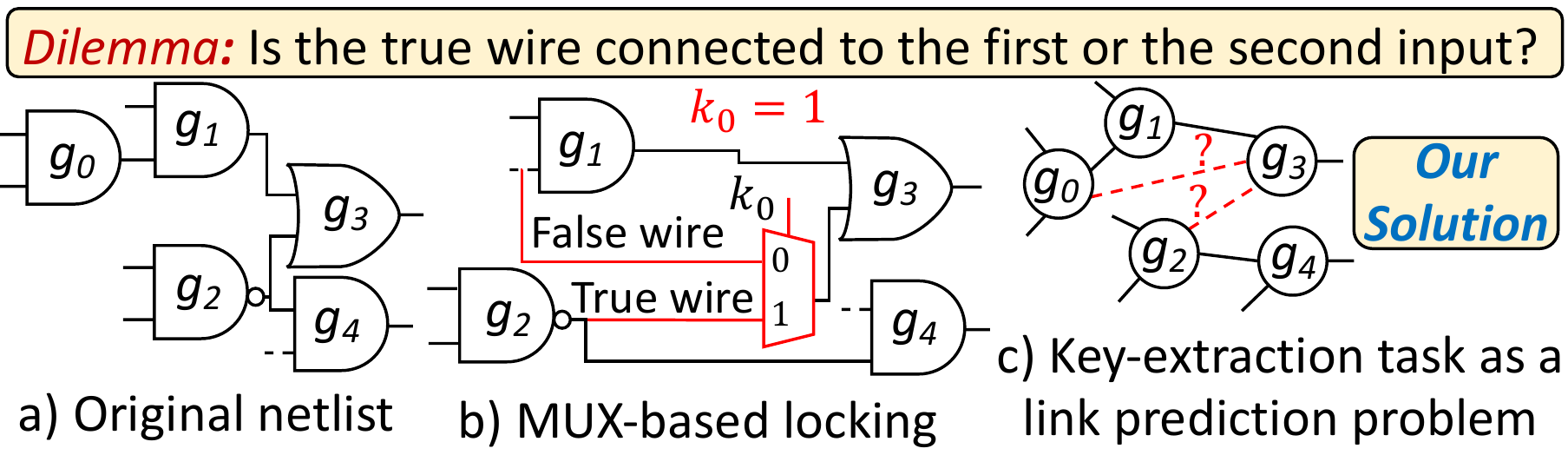}
\caption{We formulate the key-extraction task as a link prediction problem.}
\label{fig:mux_lock}
\vspace*{-2mm}
\end{figure}

\end{enumerate}

\subsection{Our Novel Concept and Contributions}

\begin{figure}[t]
\centering
\includegraphics[width=0.9\textwidth]{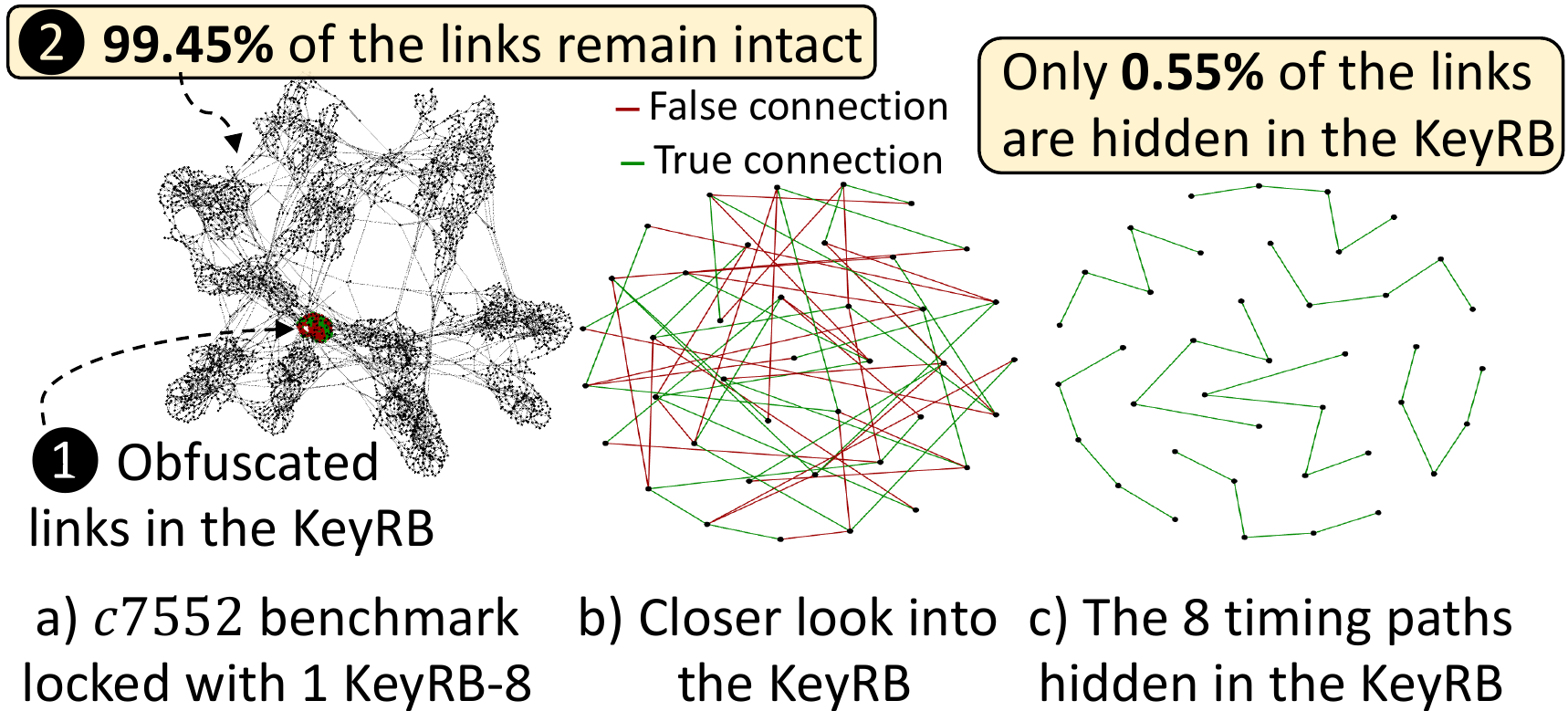}
\caption{ISCAS-85 benchmark c7552 locked using InterLock with one~KeyRB-8~\cite{InterLock}. 
Only $0.55\%$ of the links are obfuscated in the KeyRB.}
\label{fig:local}
\vspace*{-1mm}
\end{figure}

\begin{figure}[t]
\centering
\includegraphics[width=0.9\textwidth]{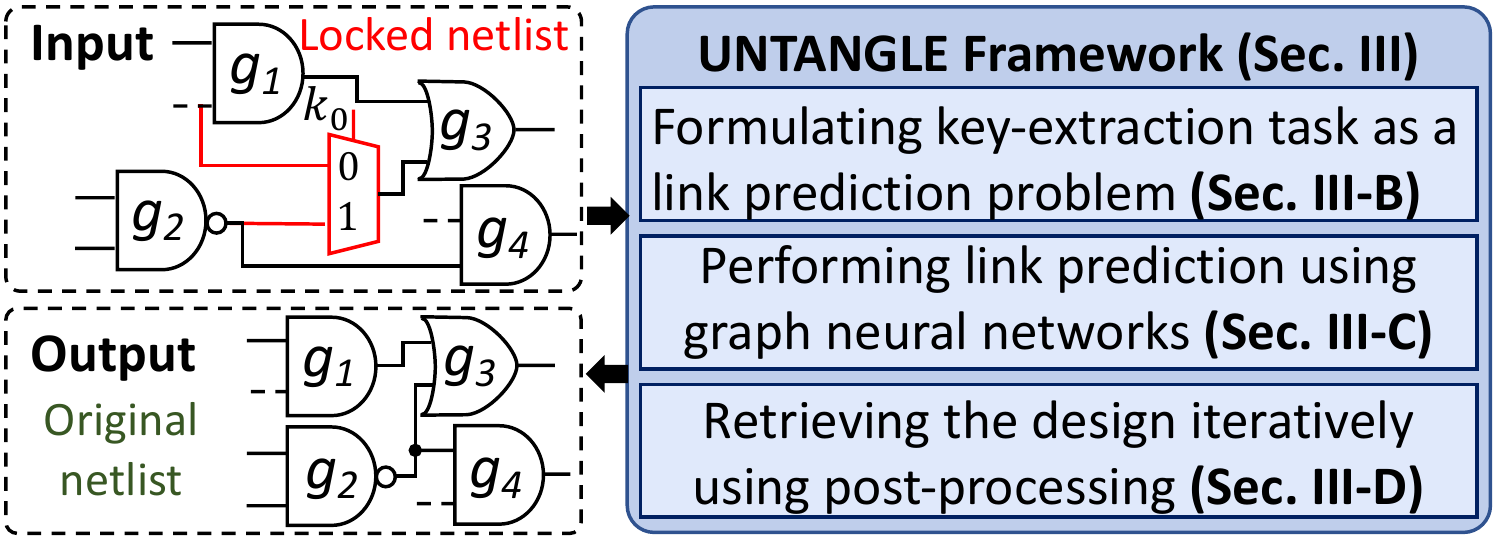}
\caption{An overview of our novel contributions.}
\label{fig:contributions}
\vspace*{-2mm}
\end{figure}

In this work, we attack routing obfuscation, focusing on the rigorous InterLock technique.
We showcase how an attacker can determine the hidden connections and gates using knowledge of the locked netlist structure (or utilizing a circuit library) \textit{without relying on an oracle}. 
The intuition behind our work is that (i)~modern ICs contain a large amount of repetition and reuse cores~\cite{saha2011soc}, and (ii)~routing obfuscation introduces limited local structural changes in the locked design, which allows the attacker to learn the remaining (intact) structure of the locked design. 
To that end, we lock the ISCAS-85 benchmark c7552 with InterLock and visualize the locked design as a graph in Fig.~\ref{fig:local}. 
The KeyRB affects a restricted portion of the design (see~\Circled{\scriptsize\textbf{1}}), leaving $99.45\%$ of the original connections untouched (see~\Circled{\scriptsize\textbf{2}}).\footnote{As the size and the number of KeyRBs increase, a larger portion of the design gets obfuscated. 
Yet, the majority of the connections remain accessible.}
Knowing which types of gates in a design are likely to be connected helps de-obfuscate the routing blocks. 
We propose {\untangle} as a generic link prediction-based attack on MUX-based locking, using graph neural networks (GNNs), as shown in Fig.~\ref{fig:mux_lock}. 
The novel contributions of this work (see~Fig.~\ref{fig:contributions}) are as follows.

\begin{enumerate}

\item \textbf{Formulating the key-extraction task as a link prediction problem:} We build a graph with edges based on the observable connections between gates outside the routing blocks. 
Then, using link prediction, we infer the links hidden in the blocks due to the routing obfuscation.

\item \textbf{Performing link prediction based on graph neural networks (GNNs):} Several heuristics exist for link prediction. 
In {\untangle}, we are interested in learning the composition of gates in the network, i.e., the graph structure (connectivity) and node features (type of gates). 
Thus, we use a GNN model that exploits the structure of the design to learn link features.
To that end, we extract local enclosing subgraphs around each considered link. 
The GNN takes in the enclosing subgraphs, utilizes the structure and gate features, and outputs vector embeddings that capture information about the target links and the composition of gates in the underlying design. 

\item \textbf{Achieving certainty of the predicted key-bits:} 
We propose a novel post-processing algorithm to unlock the design.
The algorithm examines the likelihood of each link \blue{(predicted by the GNN)} and selects only a subset of the links, which are predicted to exist with extremely high confidence. 
\blue{The selected links are then added} to the network (locked design), completing it iteratively. 
We recompute all the likelihoods
for the remaining links and the procedure continues until the network is completed, retrieving the
design with $100\%$ precision.

\end{enumerate}

\textbf{Key results:} We perform an extensive experimental evaluation of {\untangle} on selected ISCAS-85 and ITC-99 benchmarks locked using InterLock~\cite{InterLock} and random MUX-based locking. 
{\untangle} deciphers up to $100\%$ and $99.61\%$ of the key-bits, respectively, with a precision up to $100\%$. 
{\untangle} can break the locked benchmarks, which the other state-of-the-art attacks fail to unlock.
\textbf{We also open source {\untangle}~\cite{webinterface}.}
\section{Background and Related Work}
\label{sec:background}

\subsection{SAT-based Attack~\cite{Subramanyan_host_2015} and Related Countermeasures}

The SAT-based attack requires (i)~a functional IC (with the correct key embedded) acting as an ``oracle,'' \blue{and} (ii)~a locked reverse-engineered netlist. 
The attack \blue{starts by} constructing a miter using two copies of the locked netlist. 
The miter is fed to \blue{a} SAT solver to find a discriminating input pattern (DIP) for which at least two key assignments generate two different outputs. 
Subsequently, the DIP is fed to \blue{the} oracle to prune out the invalid keys.
This procedure repeats until the attack cannot determine more DIPs, resulting in the secret key. 
Researchers have developed a plethora of SAT-resistant techniques, which can be broadly categorized as follows.

\begin{enumerate}

\item \textbf{Point function-based obfuscation}~\cite{yasin_SARLock_host_2016,xie2016mitigating,yasin_CCS_2017} techniques force the SAT-based attack to rule out one incorrect key per iteration, thereby imposing an exponential number of DIPs (in terms of key-size) to unlock the design.
However, these techniques are susceptible to various structural and functional attacks~\cite{yasin_TETC_2017,yang2019stripped,sirone2020functional,alrahis2020gnnunlock}.

\item \textbf{Scan locking}~\cite{karmakar2018encrypt,wang2017secure}
techniques obfuscate the scan data,
limiting the controllability and observability of internal nets. 
Nevertheless, modeling attacks~\cite{scansat_journal,DynUnlock} have been successful in circumventing scan locking techniques.

\item \textbf{SAT-hard obfuscation}~\cite{fulllock,InterLock} techniques increase the execution time required for each SAT attack iteration by embedding key-controlled \textit{SAT-hard} instances (KeyRBs) in the design.
The KeyRBs perform routing obfuscation and are highly symmetric with different keys resulting in the same functionality (isomorphic solutions). 
These techniques are referred to as SAT-hard because symmetry is challenging for SAT solvers~\cite{fulllock}. 
Nevertheless, routing obfuscation is not sufficient to ensure security.
Modeling-based attacks can simplify the obfuscation using symmetry-breaking~\cite{sweeney_model}. 
The state-of-the-art SAT-hard InterLock technique~\cite{InterLock} performed both routing and logic obfuscation and was shown to be resistant to various state-of-the-art attacks, which we explain next.

\end{enumerate}

\subsection{InterLock--Intercorrelated Logic and Routing Locking~\cite{InterLock}}

\begin{figure}[!t]
\centering
\includegraphics[width=\textwidth]{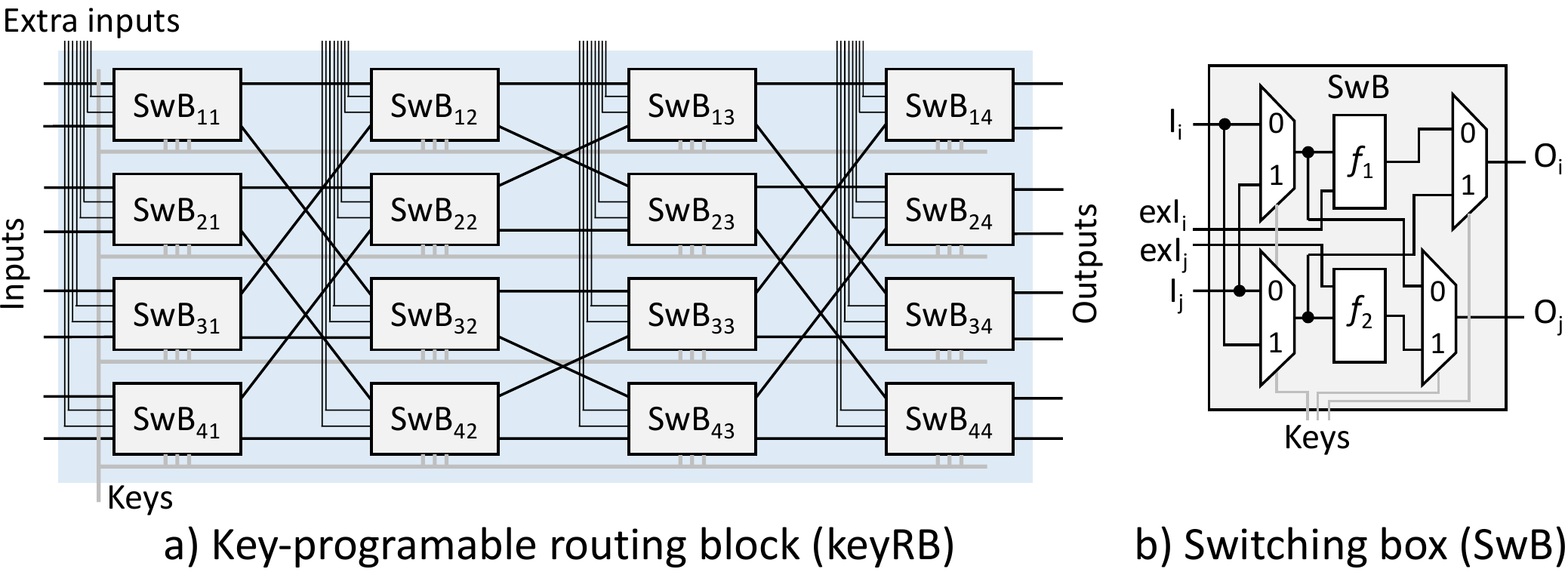}
\smallerspacecaption
\smallerspacecaption
\smallerspacecaption
\caption{KeyRB-8 in InterLock~\cite{InterLock}. $\{f_1$,$f_2$\} are 2-input gates from the circuit.\vspace{-1pt}}
\label{fig:interlock}
\end{figure}

InterLock is developed as an extension over the Full-Lock~\cite{fulllock} technique. 
In both techniques, a KeyRB is constructed using MUX-based switch boxes (SwBs), as illustrated in Fig.~\ref{fig:interlock}(a).
The KeyRB is a near non-blocking logarithmic network~\cite{network}, which has $N$ inputs, where $N$ is a power of $2$. 
The network is built using $2log_2(N)-2$ stages, where each stage consists of $N/2$ SwBs.
In Full-Lock, the SwBs are constructed using MUXes and inverters.\footnote{{\untangle} is also applicable to Full-Lock and other routing obfuscation methods. 
By (i)~re-modeling the KeyRB in Full-Lock using the all-to-all edge-encoding in~\cite{sweeney_model}, then (ii)~applying our link prediction model on the edges.} 
However, to avoid re-encoding attacks, InterLock embeds
logic gates into the keyRB using the SwBs, as depicted in Fig.~\ref{fig:interlock}(b). 
Each SwB contains four MUXes and two logic gates $\{f_1, f_2\}$. 
The MUXes are controlled by a total of three key-inputs.
The gates are constrained to be 2-input logic gates and are extracted from the original design. 
Each SwB has four inputs, $\{I_i, I_j, exI_i, exI_j\}$ and two outputs $\{O_i, O_j\}$. 
The $exI$ inputs are connected to the circuitry outside the KeyRB. 
Depending on the key, outputs $O_i$ and $O_j$ could be $\{I_i , I_j , f_1(I_i , exI_i), f_1(I_j , exI_i)\}$ and $\{I_i , I_j , f_2(Ii , exI_j), f_2(I_j , exI_j)\}$, respectively.

\begin{figure}[!t]
\centering
\includegraphics[width=\textwidth]{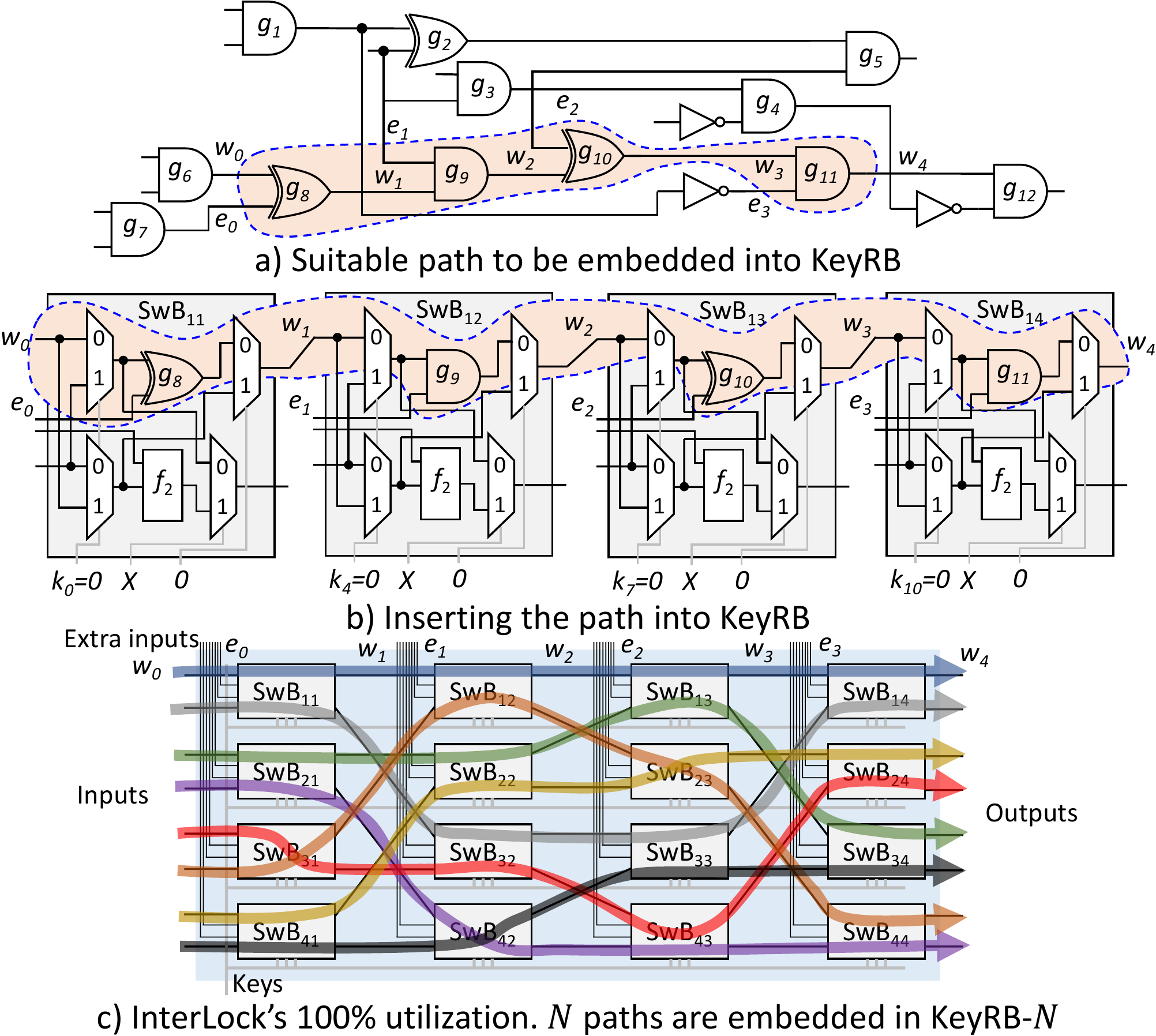}
\smallerspacecaption
\smallerspacecaption
\smallerspacecaption
\caption{Path embedding stage in InterLock (based on~\cite{InterLock}).\vspace{-1pt}}
\label{fig:interlock_path}
\end{figure}

InterLock searches for specific timing paths to incorporate into the KeyRB. 
The number of timing paths is the same as the number of inputs to the KeyRB ($N$), with a length equal to the number of stages in the block. 
We illustrate how a timing path is extracted from the original design and embedded into the network in Fig.~\ref{fig:interlock_path}.
Upon applying the correct key, the outputs of each SwB resemble the fan-outs of the gates from the original design. 
The valid key maps the original fan-ins of the $f_1$ and $f_2$ gates to the inputs of the corresponding SwBs, i.e., the outputs from the previous SwB stage.

\begin{figure}[t]
\centering
\includegraphics[width=0.7\textwidth]{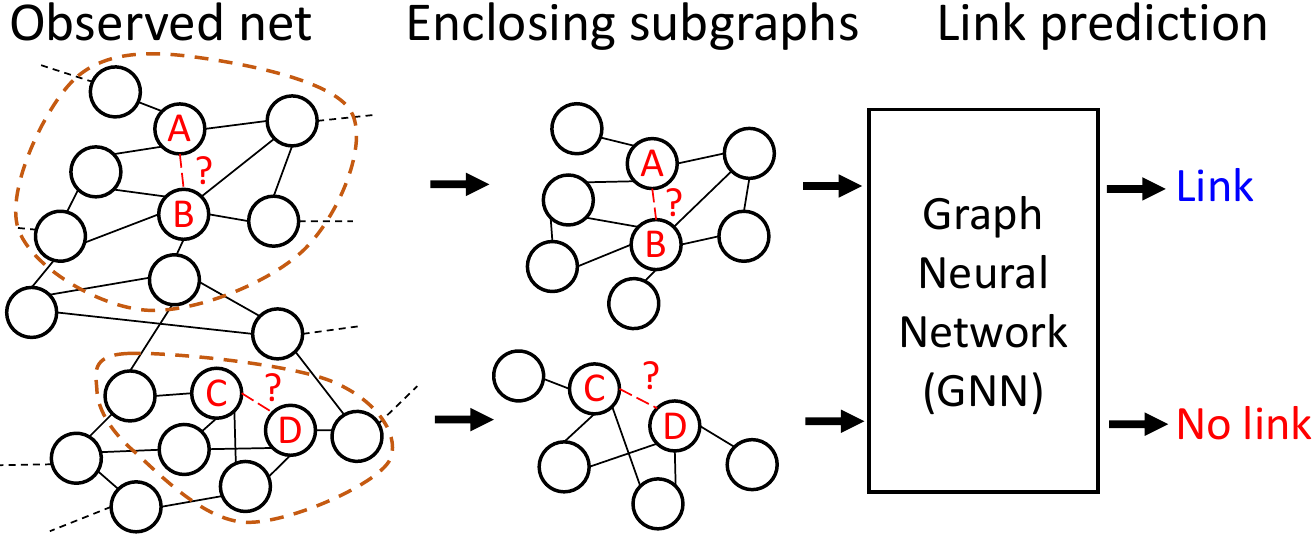}
\smallerspacecaption
\caption{Link prediction using graph neural networks (GNNs) (based on~\cite{SEAL}).}
\label{fig:seal}
\end{figure}

\subsection{Link Prediction Problem}

We infer the
timing paths in KeyRBs using link prediction. 
The underlying concept of link prediction is to estimate the likelihood of a link between two target nodes.
This estimation is governed by the structure of the observed network and the attributes of the nodes~\cite{liben2007link}. 
Link prediction has a wide variety of applications, such as protein interaction prediction~\cite{qi2006evaluation}, friend recommendation in social networks~\cite{adamic2003friends}, and drug response prediction~\cite{stanfield2017drug}.
Let $\gG=(V,E,\mX)$ be an undirected graph, where $V=\{1, 2, \ldots, n\}$ is the set of $n$ nodes, $E \subseteq V\times V$ is the set of observed edges, and $\mX \in \R^{n \times k}$ is the
matrix of node features. 
A row $\mX_{i,:}$ denotes the feature vector of node $i$ with length $k$. 
We denote the adjacency matrix of $\gG$ as $\mA \in \{0,1\}^{n\times n}$, where $\mA_{i,j}=1$ iff $(i,j)\in E$. 
Let $U$ indicates the universal set of all possible connections between vertices in the network, then $|U|=\frac{|V|(|V|-1)}{2}$.
We represent the missing links
as $T=U-E$. 
A link prediction algorithm assigns a score to all links in $T$ based on some computed heuristics. 
If the score for a link is greater than a specific threshold value, then the link is predicted to exist. 
Recently, GNNs have shown tremendous success in performing link prediction, exploiting both the structure of the graph and the associated node features to extract link features, surpassing the performance of traditional methods~\cite{SEAL}.

\subsection{Graph Neural Networks (GNNs)}

A GNN generates a vector representation (embedding) for each node in the graph such that similar nodes are placed together in the embedding space.
The embedding of a target node $v$ gets updated through message passing (neighborhood aggregation).
The features of the neighboring nodes $\gN(v)$ are accumulated to generate an aggregated representation. 
The aggregated information is then combined with the features of the target node to update its embedding. 
Consequently, after $L$ rounds of message passing, each node is aware of its features, the features of the neighboring nodes, and the structure of the graph within the $L$-hop neighborhood. 
The message passing phase is abstracted as follows, where $\mz_v^{(l)}$ indicates the embedding of node $v$ at the $l$-th round. 
\begin{align}
\vspace*{-6mm}
 \ma_v^{(l)} = \textit{AGG}^{(l)} \left( \left\lbrace \mz_u^{(l-1)} : u \in \gN(v) \right\rbrace \right) %\quad 
\vspace*{-6mm}
\end{align}
\vspace*{-5mm}
\begin{align}
\vspace*{-5mm}
 \mz_v^{(l)} = \textit{UPDATE}^{(l)} \left( \mz_v^{(l-1)}, \ma_v^{(l)} \right)
 \vspace*{-5mm}
\end{align}
GNNs mainly differ based on the choices of the \textit{AGG}$(\cdot)$ and \textit{UPDATE}$(\cdot)$ functions. 
In our work, we extract a subgraph around each target link. 
The extracted subgraphs hold information about the circuitry surrounding the link. 
Therefore, by performing graph classification, the label of the target link also becomes the label of its corresponding subgraph, as shown in Fig.~\ref{fig:seal}. 
To obtain a graph-level representation, a global pooling is applied over the node embeddings.
\section{Proposed {\untangle} Attack}
\label{sec:attack}

\begin{figure}[!t]
\centering
\includegraphics[width=0.9\textwidth]{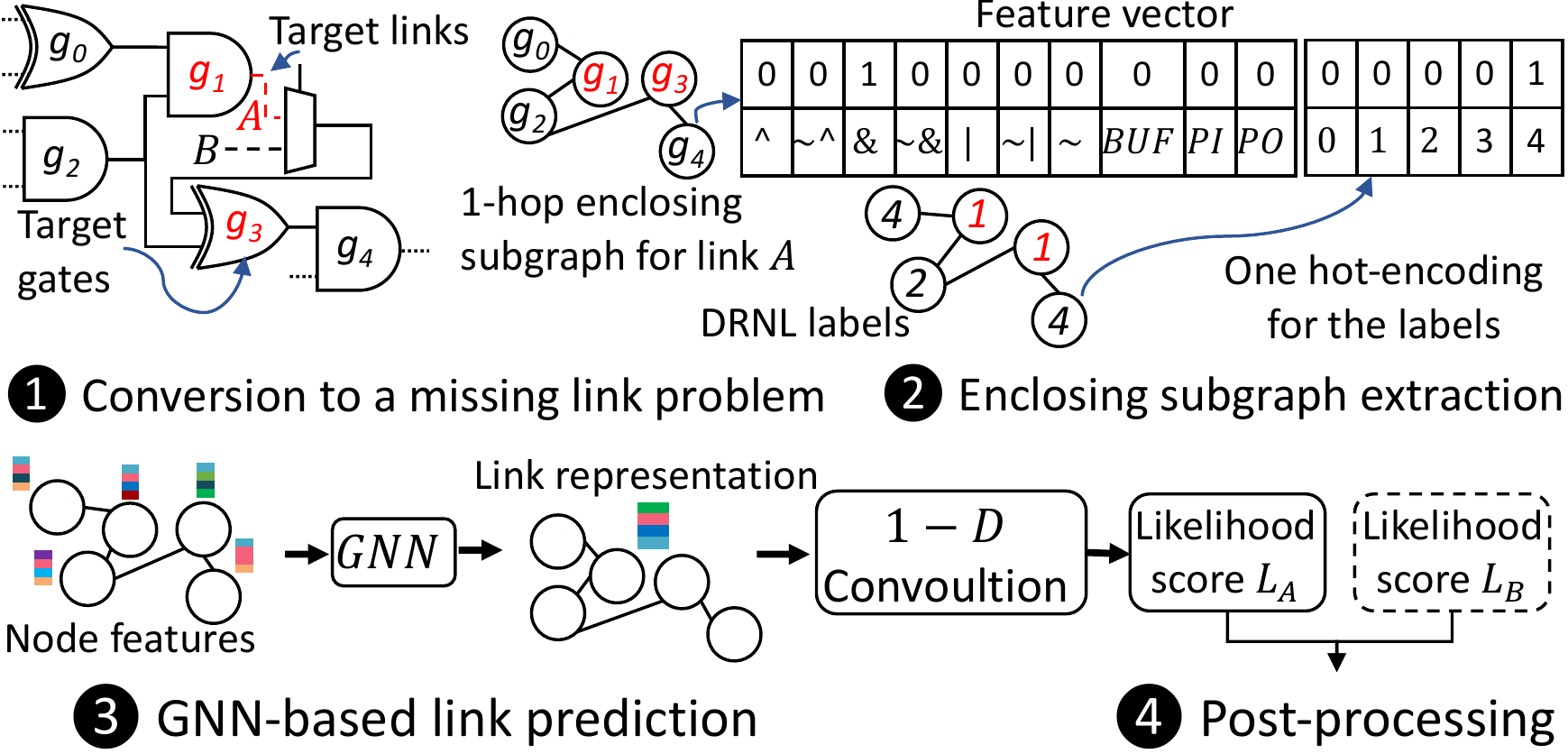}
\caption{The different steps of the proposed {\untangle} framework.}
\label{fig:steps}
\vspace*{-1mm}
\end{figure}

In this section, we provide an overview of the main steps of {\untangle} attack (Fig.~\ref{fig:steps}) and discuss the steps in detail.

\subsection{Attacker Model}

We assume an \textit{oracle-less} setting where \textit{only} the locked netlist is available.
An attacker can obtain the locked netlist by reverse-engineering the GDSII in the untrusted foundry.
The attacker can determine the location of the key-gates by tracing the key-inputs from the tamper-proof memory.

\begin{figure*}[!t]
\centering
\includegraphics[width=0.9\textwidth]{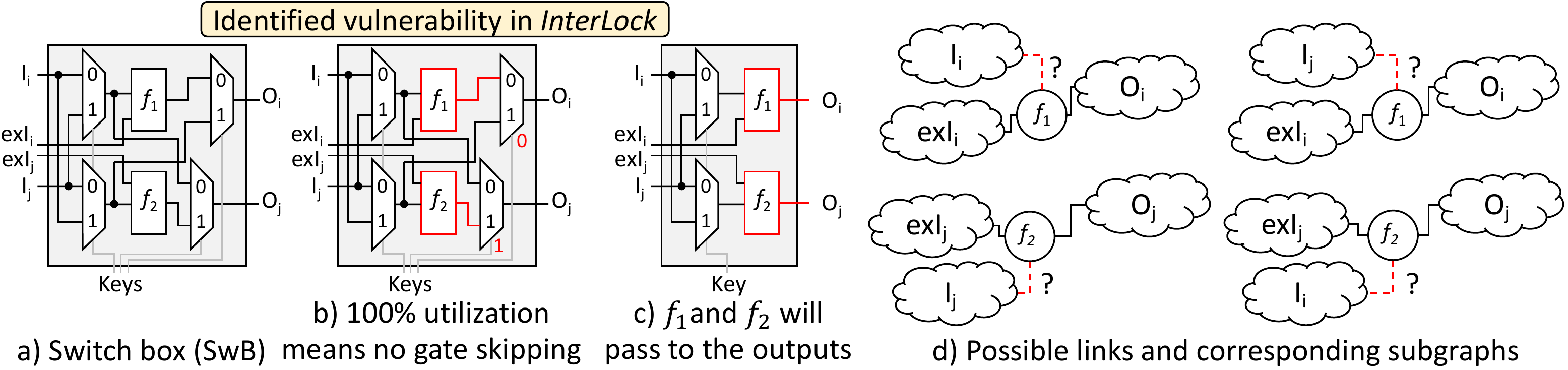}
\caption{Modeling routing de-obfuscation task as a link prediction problem.}
\label{fig:untangle}
\vspace*{-2mm}
\end{figure*}

\begin{figure*}[!t]
\centering
\includegraphics[width=0.9\textwidth]{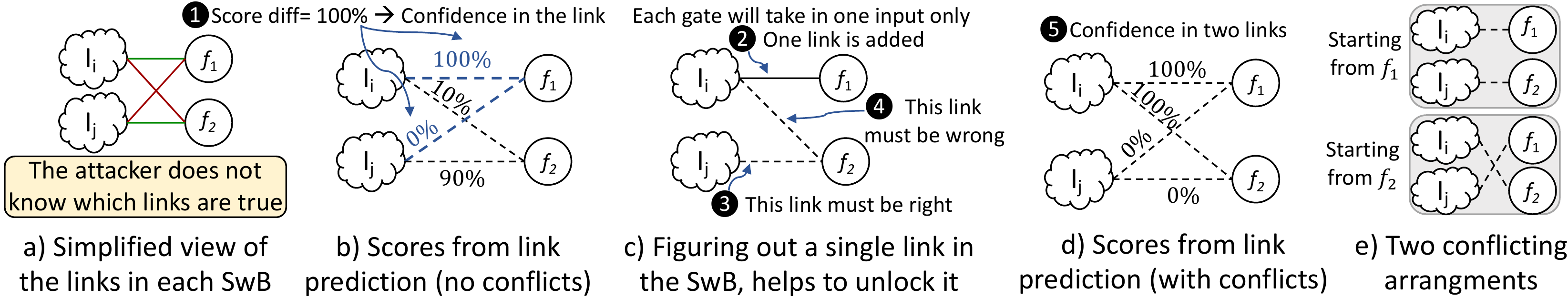}
\caption{Processing the outputs of link prediction in {\untangle} (post-processing).}
\label{fig:process}
\vspace*{-2mm}
\end{figure*}

\subsection{Formulating Key-extraction as a Link Prediction Problem}
\label{sec:links}

The resiliency of routing obfuscation comes from the complex connections introduced in the KeyRBs.
We \textit{untangle} the twisted network and consider the KeyRBs as gates with missing connections, as demonstrated in Fig.~\ref{fig:untangle}.
InterLock~\cite{InterLock} utilizes $100\%$ of the KeyRB to enhance the resilience against re-encoding attacks and \blue{to} minimize overheads. 
However, \textit{we identify a vulnerability in this implementation}, which we describe next. 
The utilization of $100\%$ indicates that each 2-input gate in a KeyRB is extracted from the original design, and therefore, cannot be skipped upon applying the correct key. 
As a result, outputs $O_i$ and $O_j$ are now restricted to $\{f_1(I_i, exI_i), f_1(I_j, exI_i)\}$, and $\{f_2(Ii, exI_j), f_2(I_j, exI_j)\}$, respectively, allowing us to infer the keys of the two independent MUXes, as illustrated in Fig.~\ref{fig:untangle}(b). 
Due to the removal of the last two MUXes (see Fig.~\ref{fig:untangle}(c)), we now consider a total of \blue{four} possible links for each SwB, as shown in Fig.~\ref{fig:untangle}(d). 
Two of the links are correct (green) and two are incorrect (red), as shown in Fig.~\ref{fig:process}(a). 
The next step is to obtain the likelihood score for each link and identify the true links.

\subsection{Link Prediction Based on Graph Neural Networks}

\subsubsection{Subgraph Extraction}

We construct an undirected graph $\gG$ based on the observable edges outside the routing block. 
Nodes in the graph map to the gates in the locked design.
We assign a one-hot encoded feature vector to each node which captures the Boolean functionality.
Additionally, the feature vector highlights if a gate has a link to a primary input (PI) or a primary output (PO). 
The length of the feature vector depends on the number of Boolean functions available in the target technology library.
We use a GNN-based platform for the link prediction task~\cite{SEAL}. 
The target nodes are grouped into set $S$. E.g., if we want to predict the likelihood of a link between nodes $u,v$, then $S=\{u,v\}$. 
Given $(S,\gG)$, an \textit{$h$-hop enclosing subgraph} $\gG_{(S,h)}$ is extracted around each pair of target nodes. 
Let $d(u,v)$ denote the shortest path distance between vertices $u$ and $v$, then $\gG_{(S,h)}$ is induced from $\gG$ by $\cup_{v\in S} \{u~|~ d(u,v) \leq h\}$.
Please refer~\Circled{\scriptsize\textbf{2}} in Fig.~\ref{fig:steps} for an example of $1$-hop subgraph extraction.

\subsubsection{Node Labeling}

We employ the double radius node labeling (DRNL) used in~\cite{SEAL} to maximize the GNN's link representation power. 
Each node in the extracted subgraph is given a label to capture its relationship with the target link. 
These labels are used as additional node attributes, which are one-hot encoded and combined with the original features, as demonstrated by~\Circled{\scriptsize\textbf{2}} in Fig.~\ref{fig:steps}. 
The target nodes are always given the unique label $1$ so that the GNN distinguishes them from the rest of the subgraph. 
Let $u$ and $v$ be the target nodes, the DRNL label $f_l(i)$ of a node $i$ is calculated as follows:
\begin{align}
\vspace*{-5mm}
f_l(i) = 1 + \text{min}(d_u, d_v) + (d / 2)[(d / 2) + (d \% 2) - 1]
\vspace*{-5mm}
\label{hashing}
\end{align}
where $d_u := d(i,u)$, $d_v := d(i,v)$, and $d := d_u + d_v$. 
In the case when $d(i,u)=\infty$ or $d(i,v)=\infty$, then $f_l(i)=0$. 
This happens if node $i$ is only connected to one of the target nodes. 
The \textit{labeling trick} is what makes graph classification suitable for link prediction. 
Instead of only learning the node features, the GNN is now aware of the target link and the relationship of the surrounding circuitry with it.

\subsubsection{GNN Model}

{\untangle} is flexible with the type of GNN to use.
We use the deep graph convolutional neural network (DGCNN)~\cite{zhang2018end}, which achieves superior results in graph classification.
A graph convolutional layer is as follows:
\begin{align}
 \vspace*{-5mm}
 \mZ^{l+1}=f(\tilde{\mD}^{-1}\tilde{\mA}\mZ^{l}\mW^{l})
 \vspace*{-5mm}
\end{align}
where $\tilde{\mA}=\mA +\mI$ adds self loops to allow self aggregation. $\tilde{\mD}$ is the diagonal degree matrix, where $\tilde{\mD}_{ii}=\sum_j \tilde{\mA_{i,j}}$, and $\mW^{l}\in\R^{k_{l} \times k_{l+1}}$ is a trainable weight matrix. $f(.)$ is an element-wise non-linear activation function. 
$\mZ^{l}\in\R^{n \times k_{l}}$ is the output embedding of layer $l-1$. 
The initial embeddings are the node features $\mZ^{0}=\mX$. The first step in the convolutional layer is $\mZ^{l}\mW^{l}$, which performs a linear feature transformation on node information, mapping the $k_{l}$ feature channels to $k_{l+1}$ channels. 
The second step aggregates the node information to neighboring vertices, including the node itself. 
Then $\tilde{\mD}$ normalizes the aggregated information to ensure a fixed feature scale.
Multiple convolutional layers can be employed to extract multi-scale sub-structure features from the network.
After $L$ layers, the output embeddings from each layer $l=1,\ldots,L$ are concatenated horizontally, to capture the graph in a single output vector $\mZ^{1:L}:=[\mZ^1, \ldots, \mZ^L]$, where $\mZ^{1:L}\in\R^{n \times \sum_{l=1}^{L} k_{l}}$. 
A \textit{sort pool layer} takes in the $n\times\sum_{l=1}^{L}k_l$ tensor $\mZ^{1:L}$ and sorts it row-wise according to $\mZ^{L}$. 
The final tensor is reshaped to $c(\sum_{l=1}^{L} k_l) \times 1$, selecting $c$ nodes to represent the graph. 
Then, the final embedding is fed to $1$-D convolutional layers with filter and step size of $\sum_{l=1}^{L}k_{l}$ to classify the graph.

\subsection{Post-processing}

Random MUX-based locking considers the location of each MUX independently. Therefore, the corresponding missing links can be processed individually.
We compare the likelihood scores of the links associated with a single MUX, and the link with the highest score gets predicted as the true wire. In the case of a tie, the corresponding key-bit will be left undeciphered. On the other hand, in InterLock, the obfuscated links are close together in the network.
Through our experiments, we conclude that completing the network iteratively enhances the performance of the attack on InterLock. 
Each link prediction step adds
more links to the network, aiding in constructing meaningful
enclosing subgraphs to predict remaining links.

We describe the {\untangle} post-processing approach for breaking InterLock in Algorithm~\ref{alg:euclid}.
As discussed in Sec.~\ref{sec:links}, four links are considered for each SwB, where two links $\{l_a, l_b\}$ are associated with each logic gate $\{f_1, f_2\}$. 
Let $T$ denote the set of all the considered links. 
The link prediction platform assigns a probability score $L_l$ for each link $l\in T$ (lines 58-64). 
In lines 22-24, the links are considered pairwise $\{l_a, l_b\}$, as shown in Fig.~\ref{fig:process}(b). 
The model looks for a pair $\{l_a, l_b\}$ having $L_a \geq up~||~L_b \geq up$ and $|L_a-L_b|\geq th$, where $th$ and $up$ are adjustable threshold and upper limit, respectively. E.g., in~\Circled{\scriptsize\textbf{1}} in Fig.~\ref{fig:process}, $up=1$ and $th=1$.
Hence, the model selects one link with high confidence (see~\Circled{\scriptsize\textbf{2}}). 
Figuring out one link in an SwB enables the model to obtain the remaining connections (see~\Circled{\scriptsize\textbf{3}} and~\Circled{\scriptsize\textbf{4}}). 
Let $C$ and $R$ denote the sets of chosen and rejected links, respectively. 
$C$ is added to the list of predicted links $P$ (line 26) and $R$ is removed from $T$ (line 27).
In case of a conflict (see~\Circled{\scriptsize\textbf{5}}), the model gets two conflicting decisions for an SwB. 
In this scenario (line 29), the post-processing restarts with an adjusted $th$ (lines 30-32). 
If the maximum $th=up=1$ is reached and there is still a conflict, an average ensemble of the network at two $h$ sizes $\{2,3\}$ (lines 35-39) is used and the likelihoods are recomputed (lines 10-12).
The GNN is not retrained for $h=3$ as using the same model (trained for $h=2$) is sufficient. 
Considering two-hop sizes together results in a robust model and prevents misclassifications. 
After selecting a set of links with high confidence, the links are removed from $T$, added to the network (lines 44-47), and link prediction is performed again.
The $th$ and $up$ values (lines 49-53) are adjusted if no links are predicted. Finally, the original design is recovered once $T$ is empty (lines 54-55).

\begin{algorithm}[!t]
\footnotesize
\caption{Pseudo-code for the proposed post-processing}
\label{alg:euclid}
\begin{algorithmic}[1] 
\footnotesize
\Require{Locked netlist graph ($\gG$), List of target links ($T$), List of obfuscated gates ($G$), and GNN model}
\Ensure{Secret key ($K$)}
\State $Done \gets FALSE$
\State $th\gets0$ \Comment{Initialize threshold}
\State $up\gets1$ \Comment{Initialize upper limit}
\State $h\gets2$ \Comment{Initialize enclosing subgraph $h$-hop distance}
\State $Ensemble\gets FALSE$
\State \textbf{Restart:}
\While{$!Done$}
\State $L\gets \{\emptyset\}$ \Comment{Likelihood scores}
\State $F\gets\{\emptyset\}$ \Comment{Scores after averaging ensemble}
 \If{$Ensemble$}
 \For{$h \in (2,3)$}
 \State $L[:,h-2]=${\getpredictions}$(h, T)$
 \EndFor
 \Else
 \State $L[:,0]=${\getpredictions}$(h, T)$
 \EndIf
 \For{$row\_number$ in $L.length()$}
 \State $F.append(mean(L[row\_number, ]))$
\EndFor
\State $L \gets F$
\State \textbf{Restart-I:}
\State $P\gets\{\emptyset\}$ \Comment{Links to add in the network}
\State $T_r\gets\{\emptyset\}$ \Comment{Links to remove from the network}
\State $G_r\gets\{\emptyset\}$ \Comment{Unlocked gates}
\For{$gate\in G$}
\If{$L_a\geq up~||~L_b\geq up$}
 \If {$|L_a-L_b|\geq th$}
 
 \If {No conflict}
 \State $P.append(C)$ \Comment{Add chosen links}
 \State $T_r.append(R)$ \Comment{rejected links}
 \State $G_r.append(gate)$
 \Else
 \If{$h=2~\&\&~th\neq up$}
 \State $th \gets th+0.1$
 \State go to Restart-I 
 \ElsIf {$Ensemble$}
 \State $Done\gets TRUE$
 \Else
 \State $Ensemble \gets TRUE$ \Comment{Activate ensemble}
 \State $th \gets 1$
 \State $up \gets 1$
 \State go to Restart \Comment{Compute the likelihoods again}
 \EndIf
 \EndIf
\EndIf
\EndIf
\EndFor
\If{!{\isempty}($P$)}
\If{$h=2$} 
\State $h\gets 3$
\State $th\gets 1$
\EndIf
%\For{$link\in P$}
\State $\gG.add(P)$ \Comment{Add predicted links to the network}
\State $T.remove(P)$ \Comment{Remove predicted links from the target links}
%\EndFor
%\For{$link\in T_r$}
\State $T.remove(T_r)$ \Comment{Remove rejected links from the target links}
%\EndFor
%\For{$gate\in G_r$}
\State $G.remove(G_r)$ 
%\EndFor
\Else
\If {$th\geq \frac{up}{2}$}
\State $th\gets th-0.1$
\Else
\State $up\gets up-0.1$
 \State $th\gets up$
\EndIf
\EndIf
\If{{\isempty}($T$)}
\State $Done \gets TRUE$
\EndIf
\EndWhile
\State $K \gets$ {\getkey}$(\gG)$ \Comment{Infer $K$ from the updated network}
\State \textbf{return} $K$\Comment{Key}
\Procedure{{\getpredictions}}{$h,T$}
\State $Temp\gets \{\emptyset\}$
\For{$link\in T$}
\State $S\gets(u,v)$ \Comment{Target gates}
\State $\gG_{(S,h)}\gets${\sample}($\gG,h,S$) \Comment{Get $\gG_{(S,h)}$ with $h$-hop sampling}
\State $Temp.append$({\gnn}($\gG_{(S,h)}$)) \Comment{Get the predictions}
 \EndFor
\State \textbf{return} $Temp$
\EndProcedure
\end{algorithmic}
\end{algorithm}

\subsection{Setup and Dataset Generation}

\subsubsection{Self-referencing Scenario}
\label{sec:self}

We train the GNN based on extracted data from the target locked design without relying on a circuit library. 
This setup does not require re-locking to be performed by the attacker. 
We use all non-obfuscated links in the locked design to create the ``positive'' training samples. 
Following the typical manner of learning-based link prediction, we randomly sample the same number of nonexistent links (unconnected node pairs) and use them as ``negative'' training data. 
We keep all the obfuscated links for testing.\footnote{By default, the target testing links do not appear in their corresponding enclosing subgraphs (because they are missing links). Hence, when extracting the samples for training, we remove the target training links from their enclosing subgraphs so that the GNN does not over-fit the training data, predicting testing links as negative because the target link does not exist~\cite{SEAL}.}

\subsubsection{Circuit Library-based Scenario}
\label{sec:lib}

The GNN is trained based on extracted data from a circuit library. 
The designs in the library are locked using the targeted locking technique. 
The library does not include the target design but includes circuits with a similar global design structure. 
Note that the Interlock technique hides specific parts of the design in the KeyRBs, leaving most of the design intact. 
We argue that a foundry with access to a library of various designs could readily identify/guess the high-level modules/functionality in the to-be-attacked design and construct a circuit library.
The training samples in this scenario additionally include the obfuscated links in the library. 
We add the true obfuscated links to the positive training samples, while the false obfuscated links are added to the negative samples.
\section{Experiments}
\label{sec:experiments}

\begin{figure}[t]
\centering
\includegraphics[width=\textwidth]{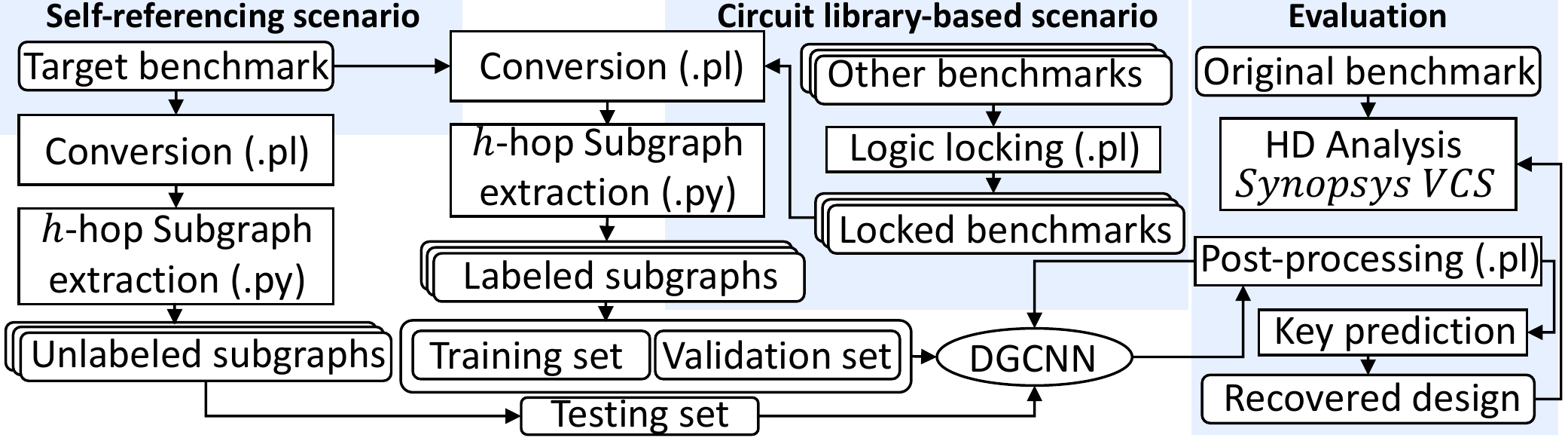}
\caption{Experimental setup and tool flow.\vspace{-1pt}}
\label{fig:setup}
\end{figure}

\subsection{Evaluation Setup, Tool Flow, and Evaluation Metrics}

We summarize the experimental setup in Fig.~\ref{fig:setup}. 
We evaluate {\untangle} on selected ISCAS-85 and ITC-99 benchmarks locked using InterLock and random MUX-based locking. 
We implement the scripts for locking and circuit to graph conversion in \textit{Perl}. 
We use the \textit{PyTorch} implementation of SEAL/DGCNN~\cite{SEAL} for link prediction, using four GNN layers with $32$, $32$, $32$, and $1$ output channels, respectively.
For the sort pooling layer, we set $c$ such that $60\%$ of the subgraphs have vertices less than $c$. 
We use two 1-D convolution layers, with $16$ and $32$ output channels and a dense layer of $128$ neurons for classification. 
Regarding the hop size $h$, it is stated that the performance saturates after $h\geq3$~\cite{SEAL}.
Thus, we train the GNN using $2$-hop subgraphs for $50$ epochs.
We use the model with the best validation performance to predict the testing links. 
We perform the experiments on an Intel(R) Xeon(R) CPU $X5680$ with $64GB$ of RAM.

\subsubsection{Dataset Generation for Random MUX-based Locking}

We insert the MUXes randomly in the designs and also randomize the selection of false wires for locking. 
We lock each ISCAS-85 benchmark
with $K:\{64, 128, 256\}$ and each ITC-99 benchmark with $K:\{256, 512\}$. 
We follow the self-referencing scenario (see Sec.~\ref{sec:self}) for random MUX-based locking. 
A feature vector of length $10$ is associated with each node. 
We further extend the feature vector by the DRNL labels.

\subsubsection{Dataset Generation for InterLock}

Similar to~\cite{InterLock}, we only lock the ITC-99 benchmarks (in \textit{BENCH} format). 
Note that as the KeyRB size increases, it embeds a larger portion of the design. 
This is why it is challenging to find suitable paths to embed in a KeyRB for small designs from the ISCAS-85 benchmark suite. 
We consider designs in \textit{BENCH} format to satisfy the restrictions of the locking technique.
The timing paths which are to be embedded in the KeyRBs must include 2-input gates only. 
We encountered challenges in meeting this requirement while handling Verilog netlists.
Hence, to ensure a fair implementation, we follow the same setup as outlined in~\cite{InterLock} and adhere to the \textit{BENCH} format.

For InterLock, we observe better performance when using the circuit library-based scenario for dataset generation (see Sec.~\ref{sec:lib}). 
The obfuscated links in InterLock are highly correlated. Therefore, when extracting random links from the remaining non-obfuscated network for training, such interference between missing links does not get captured in training. 
The designs in the library are locked using the same KeyRB instances $n$ and size $N$ as the target benchmark. 
We lock each ITC-99 benchmark with $\{1,2,3\}$ KeyRBs of sizes~$:\{8, 16\}$. 

\subsubsection{Evaluation Methods and Metrics}

In a circuit library-based scenario, {\untangle} attacks each design independently by excluding its links from training/validation. 
We report the number of correct link decisions $C$, the wrong link decisions $W$, the number of deciphered keys, and the precision.
We report the Hamming distance (HD) between the outputs of the original design and the outputs of the recovered design by {\untangle}.
For the key-bits that remain unresolved by {\untangle}, we compute the HD as follows.
For each design, we choose $100$ random keys and compare the outputs of the recovered design with the golden outputs (original design) by applying $10,000$ random input patterns using Synopsys VCS.

\subsection{Breaking InterLock~\cite{InterLock} Using {\untangle}}

In all the cases, {\untangle} achieves $100\%$ precision, implying that it always makes correct decisions when adding links to the networks. 
We report the results of attacking the benchmarks locked using 1~KeyRB-8 in Table~\ref{tab:8_8_results}. 
On average, {\untangle} deciphers $95.83\%$ of the key ($46$ out of $48$ key-bits) in an average of 2 post-processing runs (see Algorithm~\ref{alg:euclid}), leaving only two key-bits unresolved per design.\footnote{The unresolved key-bits are left for brute-force attack or SAT-based attack.}

\subsubsection{Effect of the Number of KeyRBs}

Next, we study the effect of increasing the number of KeyRBs $n$, used for locking, on the performance of {\untangle}. 
We report the results of the attack on benchmarks locked using $n:\{2,3\}$ KeyRB-8 in Table~\ref{tab:restrictions}.
\textit{The results demonstrate that {\untangle} maintains the same performance regardless of $n$.} 
{\untangle} recovers up to $97.92\%$ ($94/96$) and $98.61\%$ ($142/144$) of the key-bits for $n=2$ and $3$, respectively. 
However, with the increase in the number of missing connections, the total number of post-processing runs increases.
For example, the average number of post-processing runs required for $n=\{2,3\}$, is $\{7,10\}$. 

\subsubsection{Effect of KeyRB Size} 

Increasing the KeyRB size has a minor effect on the performance of {\untangle}. 
The average percentage of deciphered keys drops from $96.35\%$ to $89.24\%$, when 1 KeyRB-16 is used, compared to the case of KeyRB-8. 
The number of target links triples (from $32$ to $96$) and the missing links are all correlated. 
Nevertheless, {\untangle} deciphers up to $94.44\%$ ($136/144$), $93.75\%$ ($270/288$), and $96.1\%$ ($416/432$) of the key, for KeyRB-16 with $n=1$, $2$, and $3$, respectively, with $100\%$ precision (see Table~\ref{tab:restrictions}).

\begin{table}[!t]
\centering
\smallerspacecaption
\caption{\textsc{{\untangle} on benchmarks locked using InterLock with 1 keyRB-8}\vspace{-1pt}}
\label{tab:8_8_results}
\resizebox{0.9\textwidth}{!}{%
\setlength\tabcolsep{1.9pt}
\renewcommand\arraystretch{0.9}
\begin{tabular}{cccclccccccc}
\hline
\textbf{Benchmark} & \textbf{$N$} & \textbf{\begin{tabular}[c]{@{}c@{}}Attack\\ Iteration\end{tabular}} & \textbf{$th$} & \textbf{$up$} & \textbf{$h$} & \textbf{$C$} & \textbf{$W$} & \textbf{Prec.} & \textbf{\begin{tabular}[c]{@{}c@{}}Links\\ Recovered\end{tabular}} & \textbf{\begin{tabular}[c]{@{}c@{}}Links\\Left\end{tabular}} & \textbf{\begin{tabular}[c]{@{}c@{}}Total Solved\\ Key-bits\end{tabular}} \\ \hline
\multirow{2}{*}{\textbf{b22\_C}} & \multirow{13}{*}{8} & 1 & 0 & 1 & 2 & 24 & 0 & 100\% & 24 & 8 & \multirow{2}{*}{\textbf{46/48}} \\ \cline{3-11}
 & & 2 & 0.9 & 1 & 3 & 4 & 0 & 100\% & \textbf{28} & \textbf{4} & \\ \cline{1-1} \cline{3-12} 
\multirow{6}{*}{\textbf{b21\_C}} & & 1 & 0 & 1 & 2 & 14 & 0 & 100\% & 14 & 18 & \multirow{6}{*}{\textbf{46/48}} \\ \cline{3-11}
 & & 2 & 0.9 & 1 & 3 & 2 & 0 & 100\% & 16 & 16 & \\ \cline{3-11}
 & & 3 & 1 & 1 & 3 & 2 & 0 & 100\% & 18 & 14 & \\ \cline{3-11}
 & & 4 & 0.9 & 1 & 3 & 6 & 0 & 100\% & 24 & 8 & \\ \cline{3-11}
 & & 5 & 0.9 & 1 & 3 & 2 & 0 & 100\% & 26 & 6 & \\ \cline{3-11}
 & & 6 & 0.9 & 1 & 3 & 2 & 0 & 100\% & \textbf{28} & \textbf{4} & \\ \cline{1-1} \cline{3-12} 
\multirow{3}{*}{\textbf{b20\_C}} & & 1 & 0.1 & 1 & 2 & 22 & 0 & 100\% & 22 & 10 & \multirow{3}{*}{\textbf{48/48}} \\ \cline{3-11}
 & & 2 & 1 & 1 & 3 & 8 & 0 & 100\% & 30 & 2 & \\ \cline{3-11}
 & & 3 & 0 & 1 & 3 & 2 & 0 & 100\% & \textbf{32} & \textbf{0} & \\ \cline{1-1} \cline{3-12} 
\multirow{2}{*}{\textbf{b14\_C}} & & 1 & 0 & 1 & 2 & 18 & 0 & 100\% & 18 & 14 & \multirow{2}{*}{\textbf{45/48}} \\ \cline{3-11}
 & & 2 & 1 & 1 & 3 & 8 & 0 & 100\% & \textbf{26} & \textbf{6} & \\ \hline
\end{tabular}%
}
\end{table}

\begin{table}[t]
\centering
\smallerspacecaption
\caption{\textsc{{\untangle} on benchmarks locked using InterLock with $n$ KeyRBs}}
\label{tab:restrictions}
\resizebox{0.9\textwidth}{!}{%
\setlength\tabcolsep{1.9pt}
\renewcommand\arraystretch{0.9}
\begin{tabular}{ccccccccc}
\hline
\textbf{Benchmark} & \textbf{$N$} & \textbf{$n$} & \textbf{\begin{tabular}[c]{@{}c@{}}Attack\\ Iterations\end{tabular}} & \textbf{$W$} & \textbf{Prec.} & \textbf{\begin{tabular}[c]{@{}c@{}}Links\\ Recovered\end{tabular}} & \textbf{\begin{tabular}[c]{@{}c@{}}Links\\Left\end{tabular}} & \textbf{\begin{tabular}[c]{@{}c@{}}Total Solved\\ Key-bits\end{tabular}} \\ \hline
\textbf{b22\_C} & \multirow{8}{*}{8} & \multirow{4}{*}{2} & 5 & 0 & 100\% & 60 & 4 & \textbf{94/96} \\ \cline{1-1} \cline{4-9} 
\textbf{b21\_C} & & & 6 & 0 & 100\% & 56 & 8 & \textbf{92/96} \\ \cline{1-1} \cline{4-9} 
\textbf{b20\_C} & & & 8 & 0 & 100\% & 60 & 4 & \textbf{94/96} \\ \cline{1-1} \cline{4-9} 
\textbf{b14\_C} & & & 7 & 0 & 100\% & 32 & 32 & \textbf{80/96} \\ \cline{1-1} \cline{3-9} 
\textbf{b22\_C} & & \multirow{4}{*}{3} & 6 & 0 & 100\% & 92 & 4 & \textbf{142/144} \\ \cline{1-1} \cline{4-9} 
\textbf{b21\_C} & & & 7 & 0 & 100\% & 84 & 12 & \textbf{138/144} \\ \cline{1-1} \cline{4-9} 
\textbf{b20\_C} & & & 18 & 0 & 100\% & 90 & 6 & \textbf{141/144} \\ \cline{1-1} \cline{4-9} 
\textbf{b14\_C} & & & 8 & 0 & 100\% & 84 & 12 & \textbf{138/144} \\ \hline
\textbf{b22\_C} & \multirow{12}{*}{16} & \multirow{4}{*}{1} & 11 & 0 & 100\% & 68 & 28 & \textbf{130/144} \\ \cline{1-1} \cline{4-9} 
\textbf{b21\_C} & & & 16 & 0 & 100\% & 78 & 18 & \textbf{135/144} \\ \cline{1-1} \cline{4-9} 
\textbf{b20\_C} & & & 15 & 0 & 100\% & 80 & 16 & \textbf{136/144} \\ \cline{1-1} \cline{4-9} 
\textbf{b14\_C} & & & 13 & 0 & 100\% & 34 & 62 & \textbf{113/144} \\ \cline{1-1} \cline{3-9} 
\textbf{b22\_C} & & \multirow{4}{*}{2} & 12 & 0 & 100\% & 154 & 38 & \textbf{269/288} \\ \cline{1-1} \cline{4-9} 
\textbf{b21\_C} & & & 8 & 0 & 100\% & 154 & 38 & \textbf{269/288} \\ \cline{1-1} \cline{4-9} 
\textbf{b20\_C} & & & 9 & 0 & 100\% & 156 & 36 & \textbf{270/288} \\ \cline{1-1} \cline{4-9} 
\textbf{b14\_C} & & & 9 & 0 & 100\% & 152 & 40 & \textbf{268/288} \\ \cline{1-1} \cline{3-9} 
\textbf{b22\_C} & & \multirow{4}{*}{3} & 11& 0 & 100\% & 256 & 32 & \textbf{416/432} \\ \cline{1-1} \cline{4-9} 
\textbf{b21\_C} & & & 10 & 0 & 100\% & 250 & 38 & \textbf{413/432} \\ \cline{1-1} \cline{4-9} 
\textbf{b20\_C} & & &9 & 0 & 100\% & 232 & 56 & \textbf{404/432} \\ \cline{1-1} \cline{4-9} 
\textbf{b14\_C} & & & 14 & 0 & 100\% & 238 & 50 & \textbf{407/432} \\ \hline
\end{tabular}%
}
\end{table}

\subsection{Breaking Random MUX-based Locking Using {\untangle}}

For MUX-based locking, we run a single attack run because the MUXes are independent, and thus, predicting the link for a specific key-gate does not help in unlocking the rest of the key-gates. 
We report the accuracy and precision values in Table~\ref{tab:results_mux}. 
{\untangle} deciphers up to $95.31\%$ of the keys with a precision up to $100\%$, demonstrating the generic nature of our attack. 
\textit{We also launch SWEEP~\cite{alaql2019sweep} \blue{and SCOPE~\cite{SCOPE}} on these locked designs---the \blue{attacks fail} to recover any key-bit due to the existence of loops.} 
Increasing the key-size makes the observed design more incomplete due to the obfuscated (missing) connections, which impacts the performance of {\untangle}. 
For example, the accuracy drops from $89.06\%$ to $87.5\%$ when attacking c7552 locked with a key-size of $64$ and $256$, respectively. 
Note that a larger key-size also leads to higher overheads (area, power, and timing).

\begin{table}[t]
\centering
\smallerspacecaption
\caption{\textsc{{\untangle} on random MUX-based locking}\vspace{-1pt}}
\label{tab:results_mux}
\resizebox{0.85\textwidth}{!}{%
\setlength\tabcolsep{1.9pt}
\renewcommand\arraystretch{0.9}
\begin{tabular}{ccccccc}
\hline
\textbf{Benchmark} & \textbf{$K$} & \textbf{Correct keys} & \textbf{Wrong keys} & \textbf{Undeciphered keys} & \textbf{Prec.} & \textbf{Acc.} \\ \hline
\multirow{3}{*}{\textbf{c7552}} & 64 & 57 & 1 & 6 & 98.44\% & 89.06\% \\ \cline{2-7} 
 & 128 & 111 & 5 & 12& 96.09\% & 86.72\% \\ \cline{2-7} 
 & 256 & 224 & 8 & 24 & 96.88\% & 87.50\%\\ \hline
\multirow{3}{*}{\textbf{c5315}} & 64 & 61 & 0 & 3 & 100\% & 95.31\% \\ \cline{2-7} 
 & 128 & 114 & 2 & 12 & 98.44\% & 89.06\% \\ \cline{2-7} 
 & 256 & 228 & 8 &20 & 96.88\% & 89.06\%  \\\hline
\multirow{3}{*}{\textbf{c3540}} & 64 & 59 & 2 & 3 & 96.88\% & 92.19\% \\ \cline{2-7} 
 & 128 & 113 & 7 & 8 & 94.53\% & 88.28\% \\ \cline{2-7} 
 & 256 & 225 & 14 &17& 94.53\% & 87.89\%\\ \hline
\multirow{3}{*}{\textbf{c2670}} & 64 & 50 & 3 & 11& 95.31\% & 78.13\% \\ \cline{2-7} 
 & 128 & 109 & 9 &10& 92.97\% & 85.16\% \\ \cline{2-7} 
 & 256 & 212 & 16 & 28& 93.75\% & 82.81\% \\ \hline
\multirow{2}{*}{\textbf{b22\_C}} & 256 & 237 & 2 & 17& 99.22\% & 92.58\%  \\ \cline{2-7} 
 & 512 & 477 & 2 & 33 & 99.61\% & 93.16\% \\ \hline
\multirow{2}{*}{\textbf{b21\_C}} & 256 & 238 & 4 & 14 & 98.44\% & 92.97\% \\ \cline{2-7} 
 & 512 & 466 & 6 & 40 & 98.83\% & 91.02\% \\ \hline
\multirow{2}{*}{\textbf{b20\_C}} & 256 & 235 & 2 & 19 & 99.22\% & 91.80\% \\ \cline{2-7} 
 & 512 & 471 & 11 & 30 & 97.85\% & 91.99\% \\ \hline
\multirow{2}{*}{\textbf{b14\_C}} & 256 & 232 & 10 & 14 & 96.09\% & 90.63\%\\ \cline{2-7} 
 & 512 & 459 & 15 & 38 & 97.07\% & 89.65\% \\ \hline
\end{tabular}%
}
\end{table}

\subsection{HD of Designs Reconstructed by {\untangle}}

We report the HD values in Table~\ref{tab:HD}. 
{\untangle} achieves an average HD of $0.0015\%$ and $0.013\%$ when recovering designs locked with one KeyRB of size $8$ and $16$, respectively. 
The HD values indicate that {\untangle} almost obtains the exact functionality of the design \textit{without an oracle}. 
A subsequent oracle-guided attack can be carried out if an attacker desires an exact functionality (HD=$0$).
Note that since {\untangle} resolves the SAT-hard instances, the SAT-based attack will not encounter SAT-hard calls.
To that end, we launch the SAT-based attack and decipher the remaining key-bits in $8$ DIPs (on average).

\subsection{{\untangle} Run Time}

The average run time for a single post-processing run of {\untangle} on b14\_C, b20\_C, b21\_C, and b22\_C, locked using the most challenging case of $3$ KeyRB-16 is $0.32$, $0.34$, $0.44$, and $0.53$ seconds, respectively. 
The SAT-based attack runs for a day without termination on the same locked benchmarks.

\begin{table}[t]
\centering
\smallerspacecaption
\caption{\textsc{Hamming distance (HD) of designs reconstructed by {\untangle}}\vspace{-1pt}}
\label{tab:HD}
\resizebox{0.6\textwidth}{!}{%
\setlength\tabcolsep{1.9pt}
\renewcommand\arraystretch{0.9}
\begin{tabular}{ccc}
\hline
\multirow{2}{*}{\textbf{Benchmark}} & \multicolumn{2}{c}{\textbf{HD for Recovered Benchmarks}} \\ \cline{2-3} 
 & \textbf{1 KeyRB-8} & \textbf{1 KeyRB-16} \\ \hline
\textbf{b22\_C} & 0.001\% & 0.009\% \\ \hline
\textbf{b21\_C} & 0.004\% & 0.007\% \\ \hline
\textbf{b20\_C} & 0\% & 0.01\% \\ \hline
\textbf{b14\_C} & 0.001\% & 0.026\% \\ \hline
\end{tabular}%
}
\end{table}
\section{Discussion}
\label{sec:discussion}

\textbf{Comparison with Attacks:} \textit{NNgSAT}~\cite{nngsat} leverages a neural network to solve SAT-hard instances in methods such as Full-Lock~\cite{fulllock}. 
However, NNgSAT does not apply to InterLock and it requires an oracle.
The \textit{topology guided-attack}~\cite{zhang2019tga} is a dictionary-based rule learning attack that leverages the composition of gates in a circuit. 
In contrast to our approach, dictionary-based attacks cannot predict the key-bit value of a key-gate if its exact surrounding circuitry is not listed in the dictionary since it is based on exact matching. 
However, in {\untangle}, we use a GNN to learn the composition of gates, which can handle variants naturally.

\textbf{Possible Countermeasure:} {\untangle} is successful because (i)~the effects of locking are limited and local, and (ii)~the surrounding circuitry of a MUX key-gate is not obfuscated. 
A logic locking solution, which obfuscates the global structure of the design, is required. 
Large-scale MUX-based locking could be one way to get security at the expense of increasing overheads, which we plan to study in detail as part of future work. 
Furthermore, one way to improve the InterLock technique is to ensure that the KeyRBs inserted in a design are connected to each other, and the connections are obfuscated.
\section{Conclusion}
\label{sec:conclusion}

In this work, we present {\untangle}, a generic link prediction-based attack on MUX-based locking that can break the state-of-the-art SAT-hard locking technique, InterLock. 
We formulate the key-extraction task in MUX-based locking as a link prediction problem, leverage a graph neural network to learn the composition of gates in the locked netlist or a circuit library, and extract link features that assist in performing link prediction.
We demonstrate that {\untangle} can break SAT-resistant MUX-based locking (by resolving the SAT-hard instances) with precision up to $100\%$ in an oracle-less setting, which is a first in the literature. 
We believe that {\untangle} highlights the need for logic locking techniques that obfuscate the global structure of the design, as opposed to limited and local structural changes.

\newpage
\bibliography{main}
\bibliographystyle{IEEEtran} 

\ifCLASSOPTIONcaptionsoff
  \newpage
\fi

\end{document}